\documentclass[useAMS,usenatbib]{mn2e}
\usepackage{epsfig} 
 
\title[Galaxy-Galaxy Lensing by Non-Spherical Haloes I]{Galaxy-Galaxy Lensing by 
Non-Spherical Haloes I: Theoretical Considerations}
\author[Paul J. Howell and Tereasa G. Brainerd]{Paul J. Howell$^{1}$\thanks{E-mail:
phowell@bu.edu (PJH); brainerd@bu.edu (TGB)} and Tereasa G.
Brainerd$^{1}$\\
$^{1}$Boston University, Institute for Astrophysical Research, 725 Commonwealth Ave.,
Boston, MA, USA, 02215}

\begin{document}

%\date{Accepted 1988 December 15. Received 1988 December 14; in original form 1988 October 11}

%\pagerange{\pageref{firstpage}--\pageref{lastpage}} \pubyear{2002}

\maketitle
 
\label{firstpage}

\begin{abstract}
We use a series of Monte Carlo simulations to
investigate the theory of galaxy-galaxy lensing by non-spherical dark
matter haloes.  The simulations include a
careful accounting of the effects of multiple deflections on
the galaxy-galaxy lensing signal. 
In a typical observational data set where the mean tangential shear of
sources with redshifts $z_s \simeq 0.6$ is measured with respect
to the observed symmetry axes of foreground galaxies with redshifts
$z_l \simeq 0.3$, we find that the signature of anisotropic
galaxy-galaxy lensing differs substantially 
from the simple expectation that one would
have in the absence of multiple deflections.   
In general, the observed ratio of the mean tangential
shears, $\gamma^+ (\theta) /
\gamma^- (\theta)$, is strongly suppressed compared to the function that 
one would measure if the intrinsic symmetry axes of the foreground galaxies
were known.  Depending upon the characteristic masses of the lenses, the
observed ratio of the mean tangential shears may be consistent with an isotropic
signal (despite the fact that the lenses are non-spherical), or it may 
even be reversed from the expected signal (i.e., the mean tangential shear
for sources close to the observed minor axes of the lenses may exceed the
mean tangential shear for sources close to the observed major axes of the
lenses).  These effects are caused primarily by the fact that the images of
the lens galaxies have, themselves, been lensed and therefore the observed
symmetry axes of the lens galaxies differ from their intrinsic symmetry axes.
We show that the effects of lensing of the foreground galaxies on the
observed function $\gamma^+ (\theta) / \gamma^- (\theta)$ cannot be eliminated simply
by the rejection of foreground galaxies with very small image ellipticities, nor by
simply focusing the analysis on sources that are located very close to the
observed symmetry axes of the foreground galaxies.  
We conclude that any attempt to use a measurement of 
$\gamma^+ (\theta) / \gamma^- (\theta)$ to constrain the shapes of dark
matter galaxy haloes must include Monte Carlo simulations
that take multiple deflections properly into account.  
\end{abstract}
 

\section{Introduction}

Galaxy-galaxy lensing is a form of weak gravitational lensing in which
background galaxies are systematically lensed by foreground galaxies. 
Brainerd, Blandford \& Smail (1996; BBS) published the first statistically-significant
($4\sigma$) detection of this effect using a small data set that consisted
of 439 foreground galaxies, 506 background galaxies, and 3202 foreground-background
galaxy pairs.  Since this early work, galaxy-galaxy lensing has been detected with 
high precision using various data sets, most of which contain
millions of foreground-background galaxy pairs.  These high-precision
detections have allowed direct constraints to be placed on the nature of the
dark matter haloes that surround the lens galaxies, as well as on the bias between
mass and light in the universe (see, e.g., Fischer et al.\ 2000; Guzik \& Seljak 2002; 
Hoekstra, Yee \& Gladders 2004; Hoekstra et al.\ 2005; Sheldon
et al.\ 2004; Heymans et al.\ 2006; Kleinheinrich et al.\ 2006; Mandelbaum et al.\
2006ab; Mandelbaum, Seljak \& Hirata 2008;
Limousin et al.\ 2007; Parker et al.\ 2007; Natarajan et al.\ 2009; Tian et al.\ 2009).

Observations of galaxy-galaxy lensing by field galaxies have shown: [1]
at fixed luminosity, the haloes of red (early-type) galaxies are
more massive by a factor of $\sim 2$ than 
the haloes of blue (late-type) galaxies (e.g.,
Guzik \& Seljak 2002;
Kleinheinrich et al.\ 2006; Sheldon et al.\ 2004; Mandelbaum et al.\ 2006a), 
[2] the haloes of high-luminosity galaxies are more massive than
the haloes of low-luminosity galaxies (e.g., Sheldon et al.\ 2004; Mandelbaum 
et al.\ 2006a), 
and [3] the dark matter
profiles of the haloes are consistent with the spherically-averaged Navarro,
Frenk \& White (NFW) profile (Navarro, Frenk \& White 1995, 1996, 1997; 
e.g., Heymans et al.\ 2006;
Hoekstra et al.\ 2004, 2005; Kleinheinrich et al.\ 2006;
Mandelbaum et al.\ 2008).  
In other words,
observations of galaxy-galaxy lensing by field galaxies have
yielded a picture of luminous galaxies and their dark matter haloes that is
broadly consistent with the expectations of galaxy formation in the context of the 
cold dark matter (CDM) model.

Despite the popularity of the spherically-averaged NFW density
profile, CDM haloes are not
spherical.  Rather, CDM haloes are triaxial
and the degree of flattening increases with halo viral mass
(e.g.,  Warren et al.\ 1992; Jing \& Suto 2002;
Bailin \& Steinmetz 2005; Kasun \& Evrard 2005; Allgood et al.\ 2006).
In principle, galaxy-galaxy lensing should be able to provide constraints
on the shapes of the dark matter haloes of field galaxies, since a non-spherical
weak lens will produce an anisotropic shear pattern.  Consider an isolated weak
galaxy lens with a non-spherical dark matter halo (i.e., a halo
that, in projection on the sky, has an elliptical surface mass density).
For fixed source redshift and fixed angular distance
from the lens, sources that are located closer to the major axis of the lens will
experience greater shear than sources that are located closer to the minor axis 
of the lens.  If the halo of the lens can be approximated as a singular isothermal
ellipsoid with projected ellipticity $\epsilon_{\rm halo} = 0.3$,
the shear experienced
by sources nearest the minor axis of the lens will be $\sim 80$\% that of the
shear experienced by sources nearest the major axis of the lens (see, e.g., Brainerd 
\& Blandford 2002).  Although small, such an anisotropy in the 
galaxy-galaxy lensing signal should be observable
provided that, in projection on the sky, mass and light
are reasonably well aligned within the lens galaxies.

Weak lensing by galaxy clusters in the Sloan Digital Sky Survey (SDSS; e.g.,
Abazajian et al.\ 2009 and references therein)
has shown that the dark mass associated with galaxy clusters is non-spherical
and has a
projected axis ratio of $b/a = 0.48^{+0.14}_{-0.09}$ (Evans \& Bridle 2009).
The detection of non-spherical haloes by galaxy-galaxy lensing has,
however, proven to be
more problematical.  In a study of galaxy-galaxy lensing by galaxies in the
Red-Sequence Cluster Survey, Hoekstra et al.\ (2004) modeled the projected shapes
of the haloes as $\epsilon_{\rm halo} = \lambda~ \epsilon_{\rm light}$, where
$\epsilon_{\rm light}$ is the ellipticity of the image of the luminous galaxy within
the halo.  Here $\lambda=1$ indicates that
the projected shapes of the haloes are identical
to the shapes of the galaxy images, and $\lambda=0$ indicates that the haloes
are perfectly circular in projection on the sky. From their analysis,
Hoekstra et al.\ (2004) concluded that the haloes of their galaxies were
somewhat rounder than the images of the galaxies: $\lambda = 0.77^{+0.18}_{-0.21}$.
Using the same parametrization of the relationship between the ellipticities
of the haloes and the images of the galaxies, Mandelbaum et al.\ (2006b) found
$\lambda = 0.1 \pm 0.06$ for red SDSS lens galaxies and $\lambda = -0.8 \pm 0.4$ for blue 
SDSS lens galaxies.  Here the negative sign indicates an apparent anti-alignment
of mass and light for blue SDSS lens galaxies.  Finally, Parker et al.\ (2007) 
computed the galaxy-galaxy lensing signal using data from the Canada-France-Hawaii
Telescope Legacy Survey.  When Parker et al.\ (2007) averaged the
signal over all lens galaxies, they found a weak ($2\sigma$) preference for the 
haloes of the lens galaxies to be non-spherical with a projected ellipticity 
of $\sim 0.3$.
When Parker et al.\ (2007) restricted
their analysis to elliptical galaxies, the mean halo ellipticity and the significance
of the result was found to increase somewhat.

Here we construct a series of Monte Carlo simulations in order to explore the theory
of weak galaxy-galaxy lensing by non-spherical
dark matter haloes.
Using these simulations we demonstrate that, in practice, it is
challenging to {\it interpret}
the results of an observational effort to detect anisotropic galaxy-galaxy
lensing.  This is because, in general,
the observed signature of anisotropic galaxy-galaxy lensing
is strongly affected by the fact that the central, ``lens'' galaxies have, themselves
been weakly lensed.  As a result, the observed symmetry axes of the central, lens
galaxies differ from their intrinsic symmetry axes.
In our work below
we pay particular attention to the effects of multiple weak deflections on
the galaxy-galaxy lensing signal.
As was first pointed out by BBS,
galaxy-galaxy lensing is inherently a multiple deflection problem.  That is,
it is common for a source galaxy located at redshift $z_s$ to be weakly lensed
by a galaxy located at $z_{l1} < z_s$.  Oftentimes these two galaxies are 
then subsequently lensed by another 
galaxy at redshift $z_{l2} < z_{l1}$.  In other words,
the galaxy at $z_{l1}$ serves simultaneously as a lens for the galaxy at $z_s$ and
a source for the galaxy at $z_{l2}$.  In addition, the galaxy at $z_s$ is lensed
by two different foreground galaxies.  Neglecting such multiple deflections
when modeling an observed galaxy-galaxy lensing signal will give
rise to incorrect conclusions about the underlying properties of the
haloes of the lens galaxies.  For a detailed discussion of the 
frequency and relative strengths 
of multiple deflections in a deep galaxy-galaxy lensing data set, the 
reader is referred to Brainerd (2010).

Below, the haloes of the lens galaxies will be modeled as truncated
singular isothermal
ellipsoids.  This choice is motivated by two considerations.  Firstly, the
singular isothermal ellipsoid gives rise to a gravitational 
lensing shear that can be computed
analytically (e.g., Kormann, Schneider \& Bartelmann 1994).  
Secondly, at the present time the observational galaxy-galaxy lensing data are 
not of sufficiently high quality to allow one to distinguish between singular
isothermal ellipsoid haloes and those that are triaxial CDM haloes.

The outline of the paper is as follows.  In Section 2 we present the
basic theory of gravitational lensing by singular isothermal ellipsoids and
we introduce a shorthand notation that we will use throughout the paper.
In Section 3 we outline the construction of Monte Carlo simulations of
galaxy-galaxy lensing by non-spherical haloes, where the
locations and apparent magnitudes of the Monte Carlo galaxies are
taken from a large observational data set. 
In Section 4 we present the signature of galaxy-galaxy
lensing by non-spherical haloes that one should expect to obtain from a realistic
observational data set.   In Section 5 we explore the effects of galaxy-galaxy
lensing on the images of relatively nearby galaxies (i.e., galaxies that
are ordinarily be considered to be ``lenses'' but are not always considered
to be ``sources''). In Section 6 we
construct a second suite of Monte Carlo simulations in order to 
determine the effect of multiple weak deflections on observations of
anisotropic galaxy-galaxy lensing.  In Section 7 we demonstrate that
the effects of lensing of foreground galaxies on the observed signature
of anisotropic galaxy-galaxy lensing cannot be eliminated by selective 
rejection of either lens or source galaxies.  We summarize our results
and present our conclusions in Section 8.
Throughout, we will refer to the weak lensing of a background galaxy by
a single foreground galaxy as a ``deflection'', and we will adopt a flat
$\Lambda$-dominated cosmology with $H_0 = 70$~km~sec$^{-1}$~Mpc$^{-1}$,
$\Omega_{m0} = 0.3$ and $\Omega_{\Lambda 0} = 0.7$.  

\section{Truncated Singular Isothermal Ellipsoid Lenses}

Let us assume that the
dark matter haloes of large, luminous galaxies may be fairly represented as 
truncated singular
isothermal ellipsoids.  Since we are only concerned with the weak
lensing regime, the adoption of a halo model that is singular (as opposed
to a model with a finite density core) will have no effect on our 
results below.
Following Kormann et al.\ (1994), the surface mass
densities of the dark matter haloes are given by
\begin{equation}
\Sigma(\rho)  = \frac{\sigma_v^{2}\sqrt{f}}{2G}
\left(\frac{1}{\rho} - \frac{1}{\sqrt{\rho^{2} + x_{t}^{2}}}\right),
\label{surfmass}
\end{equation}
where $\sigma_v$ is the line of sight velocity dispersion,
$f$ is the
axis ratio of the
mass distribution as projected on the sky ($0 < f \le 1$), 
$x_t$ is the truncation radius, $G$ is Newton's constant,
and $\rho$ is a generalized elliptical
radius defined such that $\rho^2 = x_1^2 + f^2 x_2^2$.  Here
$x_1$ and $x_2$ are
Cartesian coordinates measured, respectively, along the minor and major
axes of the projected mass distribution of the halo.  In the limiting case of
a round lens (i.e., $f \rightarrow 1$), the total mass of the halo becomes
\begin{equation}
M_{\rm tot} = \frac{\pi \sigma_v^2 x_t}{G}.
\end{equation}

The convergence ($\kappa$) and shear ($\vec{\gamma} \equiv \gamma_1 + i \gamma_2$) 
are the characteristic
properties of a gravitational lens.  In the case of truncated singular isothermal
ellipsoid lenses, the convergence is given by
\begin{equation}
\kappa(\rho) =
\frac{\sigma_v^{2}\sqrt{f}}{2G \Sigma_c}
\left(\frac{1}{\rho} - \frac{1}{\sqrt{\rho^{2} + x_{t}^{2}}}\right),
\label{kappaeq}
\end{equation}
where $\Sigma_c \equiv  \left(\frac{4\pi G}{c^{2}}
\frac{D_{l}D_{ls}}{D_{s}}\right)^{-1}$ is the critical surface mass
density, $D_l$ is the angular diameter
distance of the lens, $D_s$ is the angular diameter distance of the source
and $D_{ls}$ is the angular diameter distance between the lens and the source.
The real and imaginary
components of the shear
can be obtained straightforwardly from equations
(63abc) of Kormann et al.\ (1994):
\begin{eqnarray}
\gamma_{1} & = & \frac{\sigma_v^{2}\sqrt{f}}{2G\Sigma_{c}}
\left[-\frac{\cos(2\varphi)}{\rho} -
\left\{f^{2}\left(x_{1}^{2}-x_{2}^{2}\right) - \left(1-f^{2}\right)x_{t}^{2}
\right\}{\mathcal P}\right] \\
\gamma_{2} & = & \frac{\sigma_v^{2}\sqrt{f}}{2G\Sigma_{c}}
\left[-\frac{\sin(2\varphi)}{\rho} -
2f^{2}x_{1}x_{2}{\mathcal P}\right]
\end{eqnarray}
where
\begin{equation}
{\mathcal P} \equiv \frac
{x_{1}^{2}+f^{4}x_{2}^{2}-
(1+f^{2})(\rho^{2}+x_{t}^{2})+
2fx_{t}\sqrt{\rho^{2}+x_{t}^{2}}}
{\sqrt{\rho^{2} + x_{t}^{2}}
\left[f^{4}r^{4}-
2f^{2}(1-f^{2})x_{t}^{2}(x_{1}^{2}-x_{2}^{2})+
(1-f^{2})^{2}x_{t}^{4}\right]}.
\label{peq}
\end{equation}
(e.g., Wright 2002).
Again,
$x_1$ and $x_2$ are Cartesian coordinates
measured along the minor and major axes of the lens, respectively.
In order to maintain consistency with the notation of Kormann et al.\ (1994),
here we have used a polar coordinate system, centred on the lens,
with radial coordinate $r \equiv \sqrt{x_1^2 + x_2^2}$ and polar angle,
$\varphi$, defined  such that $x_1 = r \cos \varphi$ and $x_2 = r\sin \varphi$.

It is clear from equations (4) through (6) that, unlike
a circularly symmetric lens for which the magnitude of the shear depends
upon the angular distance from the lens centre but not the azimuthal coordinate
of the source, the shear due to an
elliptical lens is a function of both the angular distance from the lens
centre as well as the azimuthal coordinate of the source.
At a given angular distance, $\theta$, from
the centre of an elliptical lens, the magnitude of the shear is greatest for sources
located nearest the major axis of the lens and least for sources located
nearest the minor axis of the lens.  Hence, within a given radial annulus
that is centred on the elliptical lens, 
sources whose azimuthal coordinates, $\varphi$,
place them within $\pm 45^\circ$ of the major axis of the lens  
will experience a greater mean shear than
sources whose azimuthal coordinates, $\varphi$,
place them within
$\pm 45^\circ$ of the minor axis of the lens.  As a shorthand notation, we will
refer to the magnitude of the mean shear experienced by sources whose azimuthal
coordinates place them within $\pm 45^\circ$ of the minor axis of an
elliptical lens as $\gamma^-$.  Similarly, we will refer to the magnitude
of the mean shear experienced by sources whose azimuthal coordinates place
them within $\pm 45^\circ$ of the major axis of an elliptical lens as
$\gamma^+$ (see Figure~1).

\begin{figure}
\begin{centering}
\epsfig{file=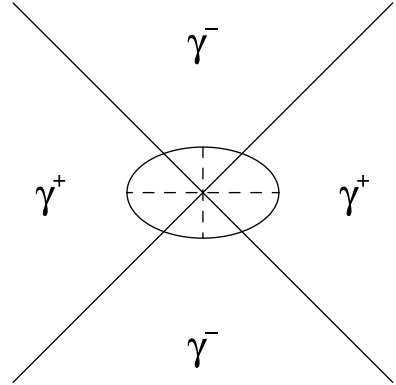,width=2in} \\
\end{centering}
\caption{Illustration of our notation, $\gamma^+$ and $\gamma^-$ (see
text).}
\end{figure}

\section{Monte Carlo Simulations of Galaxy-Galaxy Lensing
in the BTC40 Survey}

To quantify the effects of non-spherical dark matter haloes on the 
galaxy-galaxy lensing signal, we construct a series of Monte Carlo
simulations.  As a starting point for our simulations we 
use a set of  
modestly-deep, wide-field $I$-band images that were 
were generously donated to us by Emilio Falco.  The images were obtained
as part of the BTC40 survey (Monier et al.\ 2002), which 
was carried out using the Big Throughput Camera (BTC,
Tyson et al.\ 1992; Wittman et al.\ 1998) on the 4m Blanco
telescope at the Cerro-Tololo Interamerican Observatory.  The $I$-band 
exposures consist of 150 seconds per pointing, and each individual image covers
an area of order one-quarter of a square degree.  We selected
a total of 13.8~deg$^2$ of imaging data from the survey for our work, rejecting
images that were obtained during poor photometric conditions or which 
exhibited poor tracking or poor focus.  The data were calibrated, 
flat-fielded, and de-fringed as described in Monier et al.\ (2002). Object
catalogs were created from the reduced $I$-band data using the SExtractor
package (Bertin \& Arnouts 1996).

Additional details regarding the quality of the imaging, star-galaxy separation,
cosmic ray rejection, point-spread function correction,
and masking of cosmetic defects (e.g., large stellar
blooms, diffraction spikes) will be presented in a companion paper (Howell
\& Brainerd, in preparation).  In the companion paper we will also present an
analysis of the observed galaxy-galaxy lensing signal in this
data set.  For the purposes of our present study, 
we are simply interested in using the BTC40 galaxies
as the framework for a set of Monte Carlo simulations of 
galaxy-galaxy lensing by non-spherical haloes.  That is, here we will address the
following question: Given a data set like that obtained from the BTC40,
what should one expect to observe for the galaxy-galaxy lensing signal
if the dark matter haloes of the galaxies are non-spherical?
The information from the BTC40 images that we use here consists solely of the
centroids of the galaxies and their
$I$-band apparent magnitudes.
These, along with other quantities, are used as
input parameters for our Monte Carlo simulations.
Also, in order to ultimately match the data that will be
presented in our companion paper, here we use only BTC40 galaxies 
with $18 \le I_{AB} \le 22.5$.
While the completeness limit of the data is somewhat fainter than $I_{AB} = 22.5$,
in practice the BTC40 galaxies with $I_{AB} > 22.5$ are too small for accurate
shape determinations.

The observed shapes of the BTC40 galaxies have been affected by the presence
of a spatially-varying anisotropic point spread function.   Because of this,
and because of the fact that shape determinations become increasingly noisy
at faint flux levels, we do not use the observed shapes of the BTC40 galaxies
in our Monte Carlo simulations.  Instead, in order to describe the shape of the
luminous galaxy, each Monte Carlo galaxy is assigned 
an intrinsic image ellipticity, $\epsilon_{\rm in} \equiv (a-b)/(a+b)$,
that is drawn from the probability distribution derived by Ebbels (1998) from
94 archival HST field survey images:
\begin{equation}
P(\tau) = {\cal A} \tau\exp[-(\tau/0.036)^{0.54}] .
\label{edist}
\end{equation} 
Here $\tau = (a^2 - b^2)/(2ab)$, 
$\cal{A}$ is a normalising constant, and $a$ and $b$ are,
respectively, the semi-major and semi-minor axes of the intrinsic image ellipses.

We assume that
the projected shapes of the haloes of the BTC40 galaxies are
elliptical but, unlike Hoekstra et al.\ (2004) and Mandelbaum et al.\ (2006b),
we do not assume that there is a linear relationship between the shape of
the luminous galaxy and the shape of its projected
dark matter halo.  While the assumption
$\epsilon_{\rm halo} = 
\lambda \epsilon_{\rm light}$ may have some validity for elliptical galaxies, it is
definitely false for disk galaxies (which make up a substantial fraction
of the lens population).
Agustsson \&
Brainerd (2006) showed that the observed ellipticities of disk galaxies
embedded within CDM haloes were largely uncorrelated with the ellipticities
of their projected haloes (see their Figure~6).  This is due to the fact that
one always views a random projection of the dark matter halo on the sky.  
Therefore, a high inclination angle for the disk (which maximises the 
ellipticity of the luminous
galaxy image) does not, in general, correlate with a projection that maximises
the projected ellipticity of the halo.  In order to assign projected
axis ratios, $f$, to
the haloes of our Monte Carlo galaxies, then, we use the probability distribution
obtained by Agustsson \& Brainerd (2006) 
for the projected axis ratios of CDM galaxy haloes.
The halo of each galaxy in our simulations is therefore assigned a value
of $f$ that is drawn at random from this distribution (see Figure~2).

\begin{figure}
\begin{centering}
\epsfig{file=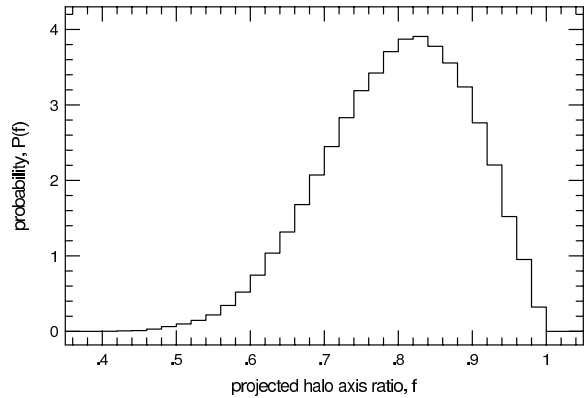,width=3in} \\
\end{centering}
\caption{Distribution of projected axis ratios, $f$, for the Monte Carlo dark
matter haloes (following
the results of Agustsson \& Brainerd 2006).
}
\end{figure}

Next we must make a choice as to how to orient the luminous galaxies within
their dark matter haloes.  
The only symmetry axes that can be used in an observational
data set to detect anisotropic galaxy-galaxy lensing are, of course, the 
symmetry axes of the luminous galaxies themselves.  If mass and
light are not reasonably well aligned within the lens galaxies, 
a detection of anisotropic 
galaxy-galaxy lensing is hopeless since we cannot directly observe the 
orientations of the symmetry axes of the dark matter haloes.   Therefore, 
in our simulations we will
assume that the intrinsic symmetry axes of the luminous galaxies and their dark matter
haloes are aligned with each other.   
This assumption maximises the degree of
anisotropy in the galaxy-galaxy lensing signal that one should expect to 
see and it presents a best case scenario for detecting the effect.

Neither spectroscopic redshifts nor photometric redshifts
are available for the BTC40 galaxies.  Therefore,
we must assign redshifts to the galaxies in order to carry out our Monte
Carlo simulations.  Following the prescriptions of BBS and Wright (2002),
we adopt a redshift distribution of the form
\begin{equation}
P(z|I_{AB}) = \frac{\beta z^2 \exp \left[ - (z/z_0)^\beta \right]}{\Gamma(3/\beta) z_0^3} .
\label{zdist}
\end{equation}
Taking $\beta = 1.5$ yields good 
agreement with the redshift surveys of LeF\`evre et al.\ (1996) and
LeF\`evre et al.\ (2004), and we then have
\begin{equation}
z_0 = 0.8 \left[0.86 + 0.15(I_{AB}-23.35) \right] .
\end{equation}

Lastly, we must assign velocity dispersions and
truncation radii to the haloes of each of the
Monte Carlo galaxies.  
To do this, we
assume that the galaxies follow a Faber-Jackson
or Tully-Fisher type of relationship and have constant mass-to-light ratio
(see, e.g., BBS).  The velocity dispersion, $\sigma_v$, of the halo of a galaxy with 
luminosity, $L$, is then given by
\begin{equation}
\frac{\sigma_v}{\sigma_v^\ast} = \left( \frac{L}{L^\ast} \right)^{1/4}
\label{veldisp}
\end{equation}
where $\sigma_v^\ast$ is the velocity dispersion of the halo of an $L^\ast$
galaxy.  The truncation radius, $x_t$, of the halo of a galaxy with 
luminosity, $L$, is given by
\begin{equation}
\frac{x_t}{x_t^\ast} = \left( \frac{L}{L^\ast} \right)^{1/2}
\label{trunc}
\end{equation}
where $x_t^\ast$ is the truncation radius of the halo of an $L^\ast$ galaxy.  The
luminosity of each Monte Carlo galaxy is obtained from its observed $I$-band
apparent magnitude
and the redshift, $z$, that was assigned to the galaxy based on equation (\ref{zdist})
above.  Accounting for the K-correction, we have
\begin{equation}
\frac{L}{L^\ast} = \left( \frac{H_0 D_l}{c} \right)
\left( 1+z \right)^{1+\alpha}  ~ 10^{0.4(22.9-I_{AB})}
\label{luminosity}
\end{equation}
where $\alpha = -\frac{d \log_{10} L_\nu}{d \nu}$ (e.g., BBS).  For
simplicity, we take $\alpha = 0.42$,
which is the mean slope of the spectral energy distribution between the Johnson $R$-band
and $B$-band from the Caltech Faint Galaxy Redshift Survey (Cohen et al.\ 1999ab).

For each Monte Carlo simulation then:
\begin{itemize}
\item a pair of characteristic parameters, $(\sigma_v^\ast, x_t^\ast)$, are 
adopted for the haloes of $L^\ast$ galaxies

\item each luminous galaxy is assigned its observed location on the 
image, as well as its observed $I$-band apparent magnitude

\item the image of each luminous galaxy is assigned an 
intrinsic shape, $\epsilon_{\rm in} \equiv (a-b)/(a+b)$,
using equation (\ref{edist}) and its dark matter halo is assigned an axis ratio, $f$,
drawn from the projected halo shapes in
Agustsson \& Brainerd (2006)

\item each luminous galaxy is assigned a random intrinsic
position angle, $\phi_{\rm in}$,
(i.e., we assume that in the absence of gravitational lensing 
the galaxy images are uncorrelated) and, since we also assume that mass and light
are aligned in projection on the sky, the projected halo
is assigned a position angle identical to the position angle 
of the unlensed luminous galaxy

\item each galaxy is assigned a redshift, $z$, using equation (\ref{zdist}),
and its luminosity relative to $L^\ast$ is obtained using equation (\ref{luminosity}) 

\item the dark matter halo
of each galaxy is assigned a velocity dispersion, $\sigma_v$, and
truncation radius, $x_t$, based upon the luminosity of the galaxy 
within the halo and the scaling 
relations of equations (\ref{veldisp}) and (\ref{trunc})

\end{itemize}
Each Monte Carlo simulation then proceeds by computing the weak lensing
shear, $\vec{\gamma}$, that is induced as light rays emanating from distant
galaxies encounter the gravitational potentials of foreground galaxies.  As 
we will see below, most of the distant galaxies with redshift $z_i$ are lensed by
numerous foreground galaxies with redshifts $z_j < z_i$.  We define the
intrinsic (unlensed) shape of each luminous Monte Carlo galaxy to be
\begin{equation}
\vec{\chi}_{\rm in} = \epsilon_{\rm in}~ e^{2i \phi_{\rm in}}
\end{equation}
where $\epsilon_{\rm in}$ in the intrinsic (unlensed) ellipticity of the galaxy
image and 
$\phi_{\rm in}$ is the intrinsic (unlensed) position angle.
Since we are dealing with the weak lensing regime, all
lensing events may be considered to be independent (e.g., Bartelmann \&
Schneider 2001) and the final image shape of each lensed galaxy is
given by
\begin{equation}
\vec{\chi}_f = 
    \vec{\chi}_{\rm in} + \Sigma_{j=1}^{N_{\rm lens}}  ~ \vec{\gamma}_j
 =  \vec{\chi}_{\rm in} + \vec{\chi}_{\rm net}
\end{equation}
where $\vec{\gamma}_j$ is the shear induced by foreground lens galaxy, $j$, and
$\vec{\chi}_{\rm net}$ is the net shear due to all foreground lenses.
The real ($\gamma_1$) and imaginary ($\gamma_2$) components of the shear
are given by equations (4) through (6) above.

Computation of the net shear for each of the galaxies due to literally {\it all}
potential
foreground lens galaxies is extremely time-consuming and, from a practical standpoint,
is unnecessary since foreground lenses that induce negligible shear
(say, $\vec{\gamma}_j \sim 10^{-9}$) can be neglected in comparison to foreground
lenses that induce substantial shear (say, $\vec{\gamma}_j > 0.005$).  From 
Brainerd (2010), we know that source galaxies with a median redshift 
$z_s = 0.96$ that have been lensed by a population of foreground galaxies with
$z_l = 0.55$ experience little shear due to lenses that are located at projected
radii $\theta > 60''$.  Scaling to the BTC40 galaxies, we find that
for $\theta > 100''$ the contribution to the net galaxy-galaxy lensing shear will
be negligible.  Hence, in our Monte Carlo simulations
we compute the net shear experienced by each BTC40 galaxy due
to all foreground galaxies that are located within a projected radius
of $100''$, and we do not include any contribution to the net shear from
lenses located at projected radii $> 100''$.

In the next section 
we will analyse the output of our Monte Carlo simulations in a manner
that is similar to the way in which an observational galaxy-galaxy lensing data set
is analysed.  In the case that neither spectroscopic nor photometric
redshifts are available for an observational data set (as is the case for
the BTC40 data), one can make only a crude distinction between ``foreground''
and ``background'' galaxies using apparent magnitudes.  From the
probability distribution above, we know that,
on average, galaxies with bright apparent
magnitudes tend to be located at lower redshifts than galaxies with faint
apparent magnitudes (though there is certainly a good deal of overlap).
If we consider galaxies with $18 \le I_{AB} \le 20$, we find
a median redshift of $z_{\rm med} = 0.29$ for our 
BTC40 sample.  If we consider
galaxies with $20 < I_{AB} \le 22.5$, we find 
a mean redshift of $z_{\rm med} = 0.61$ for our BTC40 sample.  
These magnitude cuts therefore yield a rough division of our BTC40 sample
into 38,879 ``bright''
foreground
($18 \le I_{AB} \le 20$) objects and 225,518 ``faint'' background 
($20 < I_{AB} \le 22.5$) objects.  The redshift distributions adopted for these
objects are shown Figure~3.

\begin{figure}
\begin{centering}
\epsfig{file=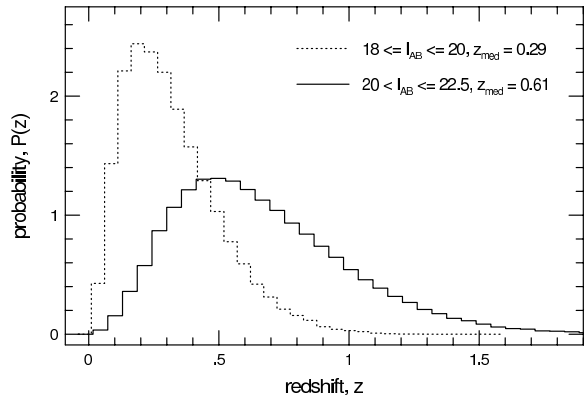,width=3in} \\
\end{centering}
\caption{Redshift distributions adopted for the ``bright'' foreground
($18 \le I_{AB} \le 20$) and ``faint'' background ($20 < I_{AB} \le 22.5$) 
galaxies in the BTC40 sample. 
}
\end{figure}

It is important to remember that in the Monte Carlo simulations, all
galaxies with redshifts $z_i$ have been lensed by all other
galaxies with redshifts $z_j < z_i$ that are located within a 
projected radius $\theta = 100''$ of the galaxy at
redshift $z_i$.  We illustrate this in Figure~4, where
we show an example of a Monte Carlo simulation.  The image corresponds to 
a single CCD frame from the BTC40 survey (0.25~deg $\times$ 0.25~deg), where
the dots indicate the locations of galaxies with magnitudes $18 \le I_{AB} \le 22.5$
and the lines indicate the directions on the sky of objects that have lensed
the galaxies.  That is, each line indicates the presence of a lens-source pair
in the data; however, for clarity of the figure we do not extend the lines to 
connect every source galaxy directly to all of its lenses.
Brown dots show the locations of bright,
foreground galaxies ($18 \le I_{AB} \le 20$)
and blue dots show the locations of faint,
background galaxies ($20 < I_{AB} \le 22.5$).
Blue lines indicate that the lens is a faint, background galaxy.
Brown lines indicate that the lens is a bright, foreground galaxy.  Therefore, 
a blue line originating from a blue dot indicates that a faint, background
galaxy has been lensed by another faint, background galaxy.  Similarly, a
brown line originating from a blue dot indicates that a faint, background
galaxy has been lensed by a bright, foreground galaxy.  Importantly, Figure~4
shows that virtually all of the bright, foreground galaxies (the brown dots)
have, themselves, been lensed multiple times.  
Most of the bright, foreground galaxies have
been lensed by other bright galaxies but they are occasionally
lensed by a faint galaxy (due to the overlapping redshift distributions of
these objects).  The majority of the lenses turn out to be faint galaxies simply
because there are $\sim 6$ times as many faint galaxies per unit area as there
are bright galaxies.

\begin{figure}
\begin{centering}
\epsfig{file=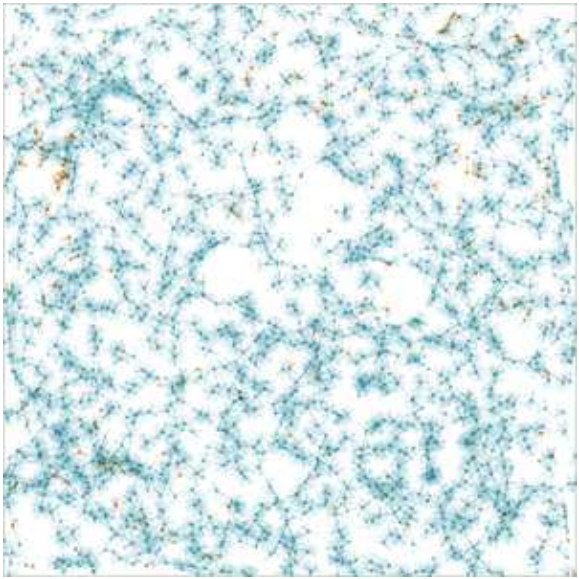,width=3.0in} \\
\end{centering}
\caption{Results of one Monte Carlo simulation for one CCD frame from
the BTC40 data.  Brown dots show the locations of bright, foreground galaxies
($18 \le I_{AB} \le 20$) and blue dots show the locations of faint, background
galaxies ($20 < I_{AB} \le 22.5$).  Lines indicate the direction on the sky of
objects that have lensed a given galaxy.  Blue lines indicate that the lens
is a faint galaxy; brown lines indicate that the lens is a bright galaxy.
Note that virtually all bright, foreground galaxies have been lensed multiple
times.
}
\end{figure}

It is clear from Figure~4 that the vast majority of the galaxies have been lensed by
more than one galaxy; that is, multiple deflections are common for 
all of the galaxies in the Monte Carlo simulations.  This statement
is true independent
of the values of the characteristic parameters adopted for the 
haloes of $L^\ast$ galaxies, $(\sigma_v^\ast,x_t^\ast)$; however, the 
relative strengths
of the individual deflections and their net effect on $\vec{\chi}_f$ for each
galaxy will, of course, be a strong function of the values of the characteristic
parameters that are adopted (see, e.g., Brainerd 2010).  

\section{Signature of Anisotropic Galaxy-Galaxy Lensing}

Here we use the output of the BTC40 Monte Carlo simulations to compute
the dependence of the mean tangential shear of the faint ($20 < I_{AB} \le 22.5$)
galaxies using the bright ($18 \le I_{AB} \le 20$) galaxies as the centres
for the calculation.  That is, we compute the signature of galaxy-galaxy
lensing in the same way as is done for an observational
 data set in which apparent
magnitude is used as the sole discriminator between ``foreground'' and ``background''
objects.  We separately compute $\gamma^{+}(\theta)$ and $\gamma^{-}(\theta)$,
and we show the results in Figures~5-7 for three different sets of characteristic
parameters that were chosen to represent the haloes of $L^\ast$ galaxies.  Since
galaxy-galaxy lensing is relatively insensitive to the radii of the dark matter
haloes (e.g., BBS; Hoekstra et al.\ 2004; Brainerd 2010), we adopt a value of
$x_t^\ast = 100~h^{-1}$~kpc for the haloes of $L^\ast$ galaxies.  In Figures~5-7
we then vary the characteristic velocity dispersion, adopting values of
$\sigma_v^\ast = 100$~km~sec$^{-1}$ (Figure~5), $\sigma_v^\ast = 150$~km~sec$^{-1}$
(Figure~6), and $\sigma_v^\ast = 200$~km~sec$^{-1}$ (Figure~7). 

\begin{figure}
\begin{centering}
\epsfig{file=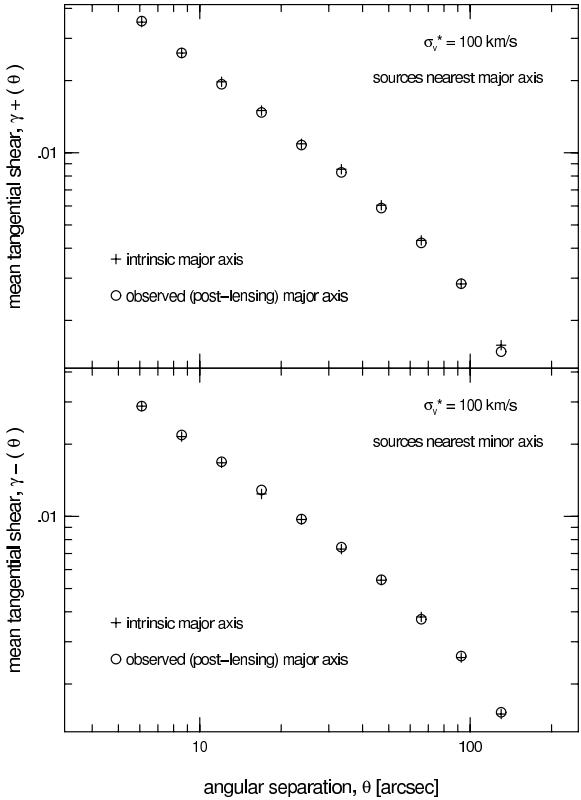,width=3in} \\
\end{centering}
\caption{Observed functions, $\gamma^+ (\theta)$ and $\gamma^- (\theta)$,
from the BTC40 Monte Carlo simulations.  
Results show the mean of 15 independent realisations of the complete BTC40
data set.
All source galaxies that
are located within $\pm 45^\circ$ of the symmetry axes of the bright,
foreground centres are used in the calculations.
Different point types indicate different definitions of the symmetry
axes of the bright, foreground centres (circles: observed symmetry 
axes after lensing, crosses: intrinsic symmetry axes). Error bars are
omitted because they are smaller than the data points.
Here the characteristic parameters
for the haloes of $L^\ast$ galaxies are $\sigma_v^\ast = 100$~km~sec$^{-1}$
and $x_t^\ast = 100~h^{-1}$~kpc.
}
\end{figure}

In order compute $\gamma^+ (\theta)$ and $\gamma^- (\theta)$ we must first
define what we mean by the symmetry axes of the bright galaxies.  Implicit in 
our definition of $\gamma^+$ and $\gamma^-$
is that the intrinsic (unlensed) symmetry axes are the symmetry axes of 
the projected dark 
matter halo (e.g., Figure~1).  However, observers are not blessed with ``dark
matter glasses'' that allow us to see the intrinsic symmetry axes.  Rather,
in an observational data set, we must take the symmetry 
axes of the bright galaxies to be their {\it observed} symmetry axes, not their
intrinsic symmetry axes. This is an important distinction 
since the observed symmetry axes of the
bright galaxies may have been 
altered due to weak lensing by foreground galaxies; 
see, e.g., Figure~4.  

In Figures~5-7 we compute $\gamma^+ (\theta)$ and $\gamma^- (\theta)$ using
both the observed symmetry axes (circles) and
the intrinsic symmetry axes (crosses) of the bright, central
galaxies.  That is, the circles indicate the
functions that we would expect to measure in an observational data set, while
the crosses indicate the functions that we would obtain if we were able to observe
the intrinsic (unlensed)
symmetry axes of the bright, central
galaxies.  In the case of very low mass lenses (Figure~5), there is relatively
little difference between the tangential shears that result from using the
observed symmetry axes of the bright centres and those that result from using
the intrinsic symmetry axes.  However, for moderate to high mass lenses 
(Figures~6 and 7), it is clear that over most scales there is
a systematic difference between the two calculations.
In particular, over most scales the observed
values of $\gamma^+ (\theta)$ in Figures~6 and 7 are systematically lower than
than the values that are obtained by using the intrinsic symmetry axes of
the bright centres.
Conversely, over most scales the observed
values of $\gamma^- (\theta)$ in Figures~6 and 7 are systematically higher than
than the values that are obtained by using the intrinsic symmetry axes of
the bright centres.

\begin{figure}
\begin{centering}
\epsfig{file=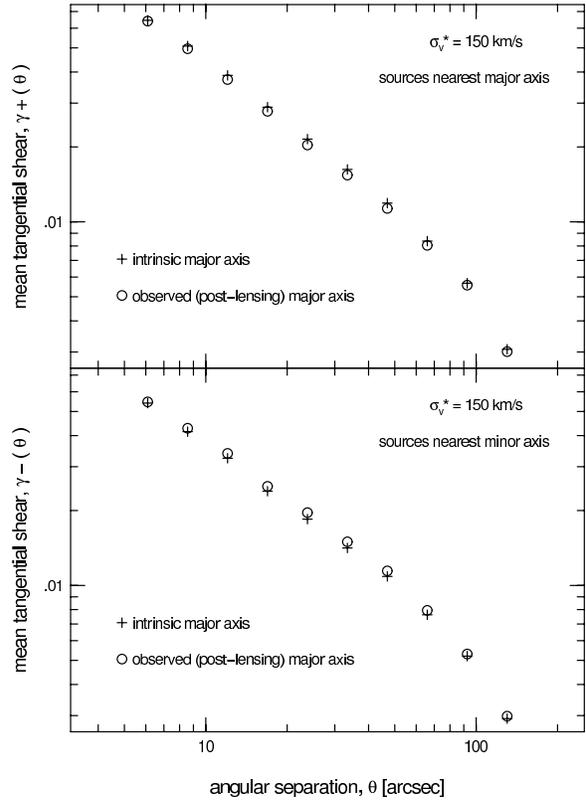,width=3in} \\
\end{centering}
\caption{Same as Figure 5, except here $\sigma_v^\ast = 150$~km~sec$^{-1}$.
}
\end{figure}

\begin{figure}
\begin{centering}
\epsfig{file=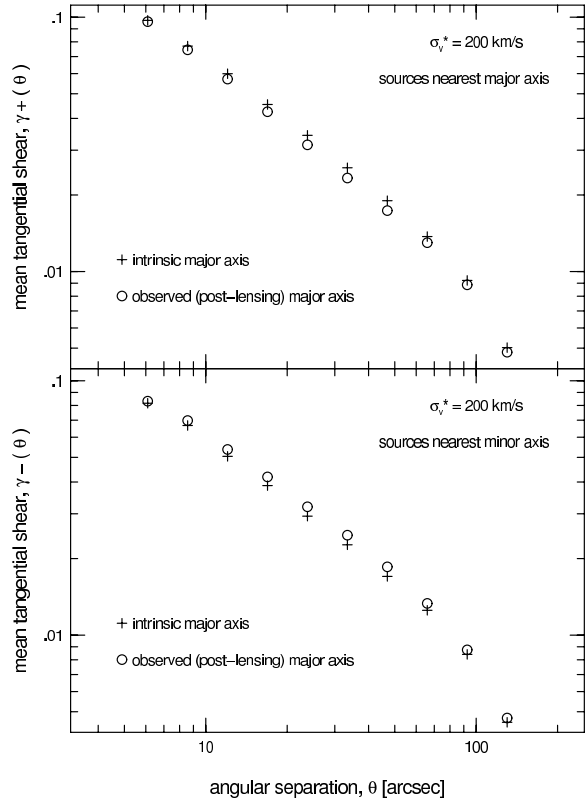,width=3in} \\
\end{centering}
\caption{Same as Figure 5, except here $\sigma_v^\ast = 200$~km~sec$^{-1}$.
}
\end{figure}

Shown in Figure~8 is the ratio of the mean tangential shears, $\gamma^+ (\theta) /
\gamma^- (\theta)$, for our three halo models.  From this figure, we see that
for low mass haloes ($\sigma_v^\ast = 100$~km~sec$^{-1}$) the ratio of the
mean tangential shears is slightly lower on average when the shear is measured
with respect to the observed symmetry axes of the bright centres than when it
is measured with respect to the intrinsic symmetry axes.  However, to within 
the error bars, the two functions formally agree.  For moderate ($\sigma_v^\ast =
150$~km~sec$^{-1}$) to high ($\sigma_v^\ast = 200$~km~sec$^{-1}$) mass lenses
there is a substantial suppression of $\gamma^+ (\theta) / \gamma^- (\theta)$
when the observed symmetry axes of the bright centres are used compared to what
one would obtain using the intrinsic symmetry axes.  In the case of $\sigma_v^\ast
 = 150$~km~sec$^{-1}$ there is little to no anisotropy apparent on scales
$\theta > 20''$.  That is, the observed signature of anisotropic galaxy-galaxy
lensing by haloes of moderate mass is largely consistent with isotropic
galaxy-galaxy lensing on scales $\theta > 20''$.
In the case of $\sigma_v^\ast = 200$~km~sec$^{-1}$, the
observed function is actually {\it reversed} from the expected function (i.e.,
$\gamma^+ (\theta) < \gamma^- (\theta)$) on scales $20'' < \theta < 70''$. 
That is, the observed signature of anisotropic galaxy-galaxy lensing by high
mass haloes could lead one to think (falsely) that mass and light are anti-aligned
within the galaxies.

\begin{figure}
\begin{centering}
\epsfig{file=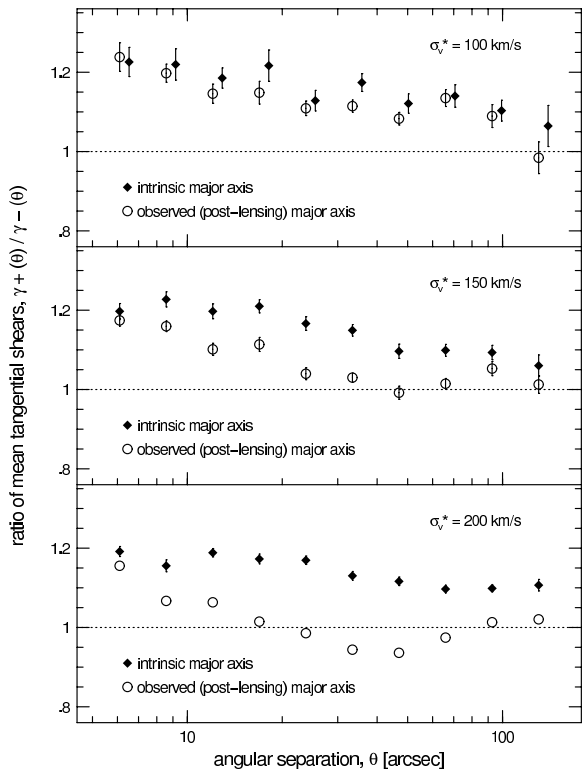,width=3in} \\
\end{centering}
\caption{Ratio of the mean tangential shears, $\gamma^+ (\theta) /
\gamma^- (\theta)$, for three halo models adopted for the BTC40
galaxies.  In all cases the truncation radius of the haloes of
$L^\ast$ galaxies is taken to be $x_t^\ast = 100~ h^{-1}$~kpc.
Here all source galaxies that are located within $\pm 45^\circ$ of
the symmetry axes of the bright, foreground centres are used
in the calculation.  Different point types indicate different
definitions of the symmetry axes of the bright, foreground centres
(circles: observed symmetry axes after lensing, 
diamonds: intrinsic symmetry axes).
Error bars are omitted when they are comparable to or smaller than
the data points.
{\it Top:} $\sigma_v^\ast = 100~$km~sec$^{-1}$.
{\it Middle:} $\sigma_v^\ast = 150~$km~sec$^{-1}$.
{\it Bottom:} $\sigma_v^\ast = 200~$km~sec$^{-1}$.
}
\end{figure}

Figure~8 demonstrates, then, that an
observation of $\gamma^+ (\theta) = \gamma^- (\theta)$ is {\it not categoric
proof that the haloes of the lenses are spherically-symmetric} since the 
haloes of our Monte Carlo galaxies are non-spherical.  In addition, an
observation of $\gamma^+ (\theta) < \gamma^- (\theta)$ is {\it not categoric
proof that mass and light are anti-aligned within the lens galaxies} since
the intrinsic symmetry axes of the luminous galaxies in the Monte Carlo
simulations were taken to be aligned with the symmetry axes of their
projected dark matter haloes.
Since we have allowed the bright, central galaxies that
we have used to compute $\gamma^{+}(\theta)$ and $\gamma^{-}(\theta)$ to
be lensed by foreground galaxies, the circles in Figure~8 show the actual
signature of galaxy-galaxy lensing by non-spherical dark matter haloes that
one should expect to see in an observational data set (i.e., a data set in which the
galaxies are all broadly distributed in redshift space).  Contrary to the
usual expectation that $\gamma^{+}(\theta)$ should exceed $\gamma^{-}(\theta)$
over a wide range of angular scales,
Figure~8 shows that this is unlikely to be the case unless the haloes of
$L^\ast$ galaxies have particularly low characteristic velocity dispersions
($\sigma_v^\ast = 100$~km~sec$^{-1}$).   In Section 6
we will demonstrate that the results shown in Figure~8 are
caused primarily by the fact the observed
symmetry axes of the bright, foreground centres
have been altered from their
intrinsic symmetry axes by weak lensing.

Figures~9 and 10 show schematic illustrations of what can occur in
the situation that a given lens-source pair is, itself, lensed by a foreground
mass.  Consider an intrinsically circular
source galaxy that is located near the major axis of an elliptical lens galaxy.  
That is, the lens-source pair is in what one might call the ``$\gamma^{+}$
configuration''.  After being sheared by the elliptical lens, the image of
the source galaxy is an ellipse with the major axis of its image 
oriented tangentially with respect to the major axis of the elliptical lens.  This is 
illustrated by the blue ellipse in the top and bottom panels
of Figure~9.  Now consider the effect on the
image of the intrinsically circular source if a large mass (i.e., another
galaxy) is placed in the foreground of the original lens-source pair.  If
the additional foreground mass is located along the line that connects the centroids
of the original lens-source pair, the net result for the image of the 
intrinsically-circular source is that it becomes {\it more elliptical} than
if it had been lensed only once.  This is illustrated by the red
ellipse in the top panel
of Figure~9.  Now consider the effect if the foreground mass is placed
such that its location is tangential to the line that connects the centroids
of the original lens-source pair.  The net result for the image of
the intrinsically-circular source is that it will be {\it less elliptical}
than if it had been lensed only once.  This is illustrated by the
red ellipse in the
bottom panel of Figure~9.  We see, therefore, that the inclusion of a second lens
may either increase or decrease the net ellipticity of our distant, circular
source over what we would have naively expected in the single-deflection 
case.

\begin{figure}
\begin{centering}
\epsfig{file=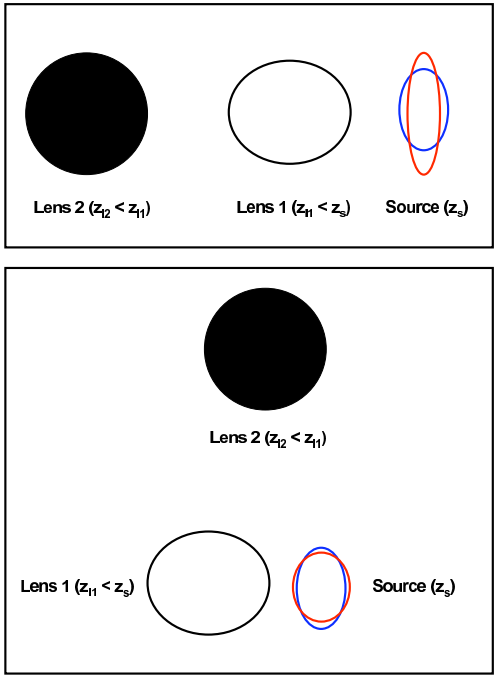,width=3in} \\
\end{centering}
\vskip 0.5cm
\caption{Schematic illustration of the net effect on the image of
a distant, intrinsically-circular source galaxy due to an elliptical 
lens with redshift $z_{l1} < z_s$ and an additional, foreground
lens with redshift $z_{l2} < z_{l1}$.  The black ellipse 
indicates the true shape and orientation of the elliptical lens.
In both panels the elliptical lens and the source are intrinsically
in the ``$\gamma^+$''
configuration.  Blue ellipses: shape of
the source after lensing solely by the elliptical lens.  Red ellipses:
shape of the source after being lensed by the elliptical lens
at $z_{l1}$ and the additional foreground lens at $z_{l2}$.
When the foreground lens is located along the direction vector that
connects the centroids of the elliptical lens and the circular source,
the final ellipticity of the image of the source is increased
compared to what it would have been if the source had been
lensed solely by the elliptical lens.
When the foreground lens is located tangential to the direction vector
that connects the centroids of the elliptical lens and the circular
source, the final ellipticity of the image of the source is
reduced compared to what it would have been if the source had
been lensed solely by the elliptical lens. 
}
\end{figure}

Next let us consider the fact that not only will the introduction of an 
additional foreground mass alter the image
of the distant, circular source galaxy, it
will also alter the image of the 
original elliptical lens galaxy.  Again, consider
the original lens-source pair to be in the $\gamma^{+}$ configuration.  If
the foreground mass is placed along the line that connects the centroids
of the original lens-source pair, the image of the elliptical lens becomes
rounder than its intrinsic shape 
(i.e., since it is distorted tangentially with respect to the 
location of the foreground mass).  In some finite number of cases where the 
intrinsic ellipticity of the image of the elliptical lens is very small
and the shear due to the foreground lens is large, 
the post-lensing image of the elliptical lens may even have
its observed symmetry axes reversed from its intrinsic symmetry axes.  As
a result, the distant, intrinsically circular source galaxy would appear to
be in the $\gamma^{-}$ configuration when, in fact, it is in the $\gamma^{+}$
configuration (e.g., top panel of Figure~10).  
Should the situation illustrated in the top panel of
Figure~10 occur, the mean tangential shear that one would
obtain for sources
located close to the {\it observed minor axes} of the elliptical lens will be 
greater than it ought to be.    That is, the observed value of
$\gamma^{-}$ is boosted by the incorrect inclusion of sources that would have
properly gone into the calculation of $\gamma^{+}$ if one had known the orientation
of the intrinsic symmetry axes of the elliptical lens.  Of course, this
also results in the observed value of $\gamma^{+}$ being reduced compared to
its true value because the observed
lens-source configuration has been ``misclassified''
compared to its intrinsic configuration.

\begin{figure}
\begin{centering}
\epsfig{file=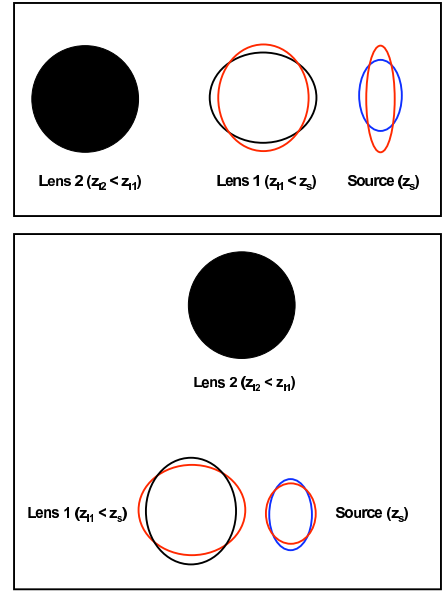,width=3in} \\
\end{centering}
\vskip 0.5cm
\caption{Schematic illustration of the net effect on 
(i) the image of
a distant, intrinsically-circular source galaxy due to an elliptical 
lens with redshift $z_{l1} < z_s$ and an additional, foreground
lens with redshift $z_{l2} < z_{l1}$, and (ii) 
the image of the elliptical lens after being lensed by the foreground
lens at $z_{l2}$.  The black ellipse 
indicates the true shape and orientation of the elliptical lens
in both panels.
Top: elliptical lens and source are intrinsically
in the ``$\gamma^+$'' configuration.
Bottom: elliptical lens and source are intrinsically
in the ``$\gamma^-$'' configuration.
Blue ellipses: shape of
the image of the 
source after lensing solely by the elliptical lens.  Red ellipses:
shape of the image of the elliptical lens after being lensed by the foreground
lens at $z_{l2}$, as well as the 
shape of the image of the source after being lensed by the elliptical lens
at $z_{l1}$ and the additional foreground lens at $z_{l2}$.
This figure illustrates that
in the limit of small intrinsic ellipticities, the observed 
symmetry axes of the elliptical lens may be reversed from its 
intrinsic symmetry axes.  This results in a misclassification of the 
configuration ($\gamma^+$ or $\gamma^-$) of the elliptical lens-source pair,
which incorrectly enhances measurements of $\gamma^-$ and incorrectly
suppresses measurements of $\gamma^+$.
}
\end{figure}

Next consider placing the foreground mass along a line that is tangential
to the line that connects the centroids of the original lens-source pair.
If the original lens-source pair is in the $\gamma^{+}$ configuration, the
image of the elliptical lens will have an increased ellipticity after being
lensed by the foreground mass and it will not undergo a reversal of its
symmetry axes (i.e., the lens-source pair remains in the $\gamma^+$
configuration), but the final image of the source will be less elliptical
than it would have been in the absence of the foreground lens.  In the case that
the lens-source pair is intrinsically in the $\gamma^{-}$ configuration, however,
a finite number of lenses with small intrinsic ellipticities may have their
symmetry axes reversed by lensing due to the foreground mass (e.g.,
bottom panel of Figure~10).  Hence, the
original lens-source pair would appear to be in the $\gamma^{+}$ configuration,
when, in fact, it is actually in the $\gamma^{-}$ configuration.  In this
case, the intrinsically circular source is incorrectly put into the calculation
of $\gamma^{+}$, and its net shear will be rather small because: (i) it is, in reality,
located near the minor axis of the elliptical lens and (ii) after being lensed
by the additional foreground mass, its image will be rounder than if it had been lensed
solely by the original elliptical lens.  Both of these conspire to reduce the
observed value of $\gamma^{+}$ compared to its true value.  

Based upon relatively simple reasoning from Figures~9 and 10, we 
may therefore expect that multiple deflections (i.e., the presence of more than one
foreground lens) could lead to a suppression of $\gamma^+ (\theta) / \gamma^- (\theta)$
compared to what would would obtain if one knew the intrinsic symmetry axes
of the bright centres. One might hope that in a sufficiently large data set
the effects of foreground lenses on the original elliptical lens-source pair
would cancel each other out.  However, this is not necessarily going to be the case.
Galaxies span a broad range of redshifts; hence, at
fixed angular separation from a source, $\theta$, two foreground lens
galaxies with
identical gravitational potentials will have different lensing strengths 
because they are located at different physical distances from the source.
In the following sections we will
explore the effect of galaxy-galaxy lensing
on the images of the bright, foreground objects that are used as centres
for the computation of the 
mean tangential shear of the faint, background objects.  In addition, we will
explore the effect of weak lensing of the central galaxies on the measured
values of $\gamma^+(\theta)$ and $\gamma^-(\theta)$.

\section{Effect of Galaxy-Galaxy Lensing on Foreground Galaxy Images}

All of our Monte Carlo 
galaxies with redshifts $z_i$ have been lensed by all other Monte Carlo galaxies 
with redshifts
$z_j < z_i$ that are found within a radius of $100''$ of the galaxy at $z_i$.
Since the redshift distribution of the galaxies is broad, this means that many
of the bright centres (i.e., those BTC40 galaxies with $18\le I_{AB} \le 20$) correspond
to galaxies that have, themselves, been lensed (e.g., Figure~4).  
At a given angular separation from a weak galaxy lens, galaxy-galaxy lensing 
may be considered to be a scalar shear.  That is, although a source galaxy has
a finite size on the sky, the shear due to a foreground weak galaxy lens is
effectively constant across the image of the source.  
Shown in Figure~11 is an illustration
of the transformation of an ellipse (i.e., the intrinsic shape of a galaxy) 
due to a scalar shear.  The important things to note from this figure are that
a scalar shear applied to an ellipse results in a change in the ellipticity
as well as a change in the position angle.  In addition, 
the original major and minor axes of the ellipse are no longer orthogonal
after the transformation.

\begin{figure}
\begin{centering}
\epsfig{file=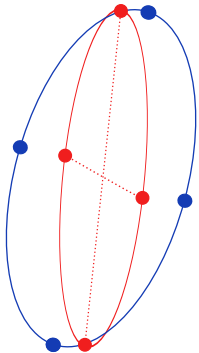,width=1in} \\
\end{centering}
\caption{Transformation of an ellipse by a scalar shear.  Blue: original
ellipse.  Red: transformed ellipse.  Note that not 
only are the ellipticity and position angle of the ellipse
altered, the original major 
and minor axes of the ellipse are no longer orthogonal.
}
\end{figure}

In the limit that the intrinsic ellipticity, $\epsilon_{\rm in} \equiv (a-b)/(a+b)$,
of an ellipse is large compared to the applied shear, $\epsilon_{\rm in} >> \gamma/2$,
the transformation of the ellipse due to a scalar shear can be obtained
straightforwardly from equations
(39) and (40) of Surpi \& Harari (1999).  Let the ellipse have
intrinsic axis ratio $f = b/a$ and intrinsic 
position angle $\psi$.  If we then take $\phi$ to be
the angle between the major axis
of the unlensed source ellipse and the vector that connects the centroids of the
lens and the source ellipse, 
the resulting change in position angle of the source ellipse is
\begin{equation}
\Delta \psi \simeq \gamma \left( \frac{1 - f^2}{1 + f^2} \right)
\sin 2(\phi-\psi) 
\end{equation}
and the square of the axis ratio of the transformed source ellipse is given by
\begin{equation}
\left( f' \right)^2 = \frac{f^2 - 2\gamma f^2 \cos 2(\phi-\psi)}{1 +
2 \gamma f^2 \cos 2(\phi-\psi)} .
\end{equation}
Figure~12 shows the resulting change in position angle and ellipticity for
ellipses with intrinsic ellipticity $0.02 \le \epsilon_{\rm in} \le 0.3$ due
to a 1\% scalar shear ($\gamma = 0.01$).  Unsurprisingly, the smaller is
$\epsilon_{\rm in}$, the greater is the change in position angle, and the 
maximum change in the position angle occurs for $(\phi-\psi) = 45^\circ$.
In addition, the smaller is $\epsilon_{\rm in}$, the greater is the change 
in ellipticity. The maximum change in ellipticity occurs for the two 
extreme conditions: $(\phi - \psi) \sim 0^\circ$, resulting in an ellipse 
that is rounder than its intrinsic shape, and $(\phi - \psi) \sim 90^\circ$,
resulting in an ellipse that is flatter than its intrinsic shape.

\begin{figure}
\begin{centering}
\epsfig{file=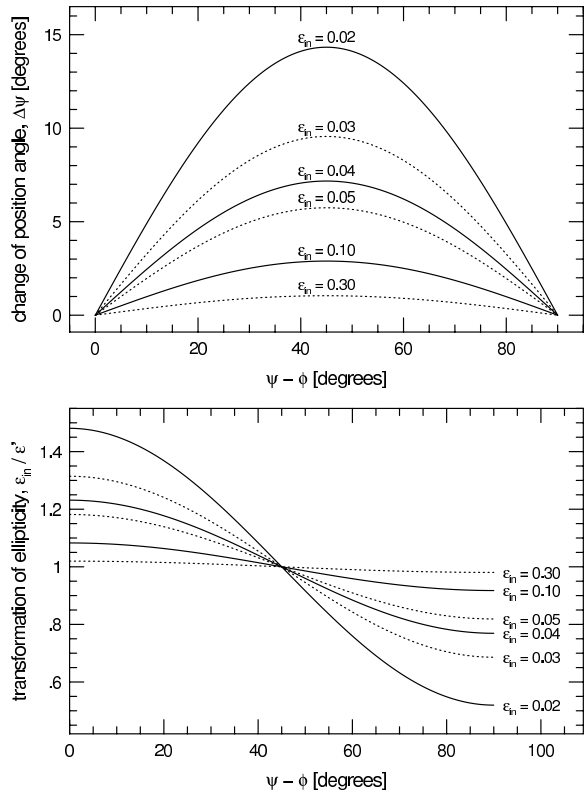,width=3in} \\
\end{centering}
\caption{Transformation of an ellipse of intrinsic ellipticity, $\epsilon_{\rm in}$,
due to a scalar shear of magnitude $\gamma = 0.01$.  Prior to being sheared,
the position angle of the ellipse is $\psi$.  The quantity $\phi$ is the angle
between the major axis of the unsheared source ellipse and the direction
vector that connects the centroids of the lens and source.
Top: change in position angle of the ellipse.  Bottom: ratio of intrinsic and
transformed ellipticities.
}
\end{figure}

\section{Effect of  Multiple Weak Deflections on $\gamma^+(\theta)$
and $\gamma^-(\theta)$}

In the previous section we demonstrated that, in the limit of small intrinsic
ellipticities, the images of the galaxies that
correspond to the ``bright centres'' in our galaxy-galaxy lensing
analysis could be significantly affected by weak lensing
by foreground galaxies under the right circumstances.  
That is, multiple weak deflections (in which a
lens-source pair is subsequently lensed by one or more foreground galaxies)
will affect not only the image of the original source, but they will also 
affect the image
of the original lens.
In this section we investigate the effect of such multiple weak
deflections 
on measurements of $\gamma^+(\theta)$ and $\gamma^-(\theta)$.
Here we wish to separate the effects of weak lensing due to non-spherical dark
matter haloes from the effects of subsequent multiple weak deflections due
to the haloes of foreground galaxies.  
To address this, we construct a set of constrained Monte
Carlo simulations in which a lens-source pair that contains an elliptical
lens is, itself, weakly-lensed by additional galaxies whose dark matter 
haloes are spherically-symmetric.  

This second set of Monte Carlo simulations is constructed as follows.  An
elliptical lens of fixed mass axis ratio, $f$, is placed at the origin of
the coordinate system.  Here $f$ is the axis ratio of the projected
dark matter halo.  The elliptical lens is assigned a random position angle and
a redshift, $z_{el}$, drawn from the
redshift distribution adopted for the BTC40 objects with apparent magnitudes
$18 \le I_{AB} \le 20$.  The elliptical lens is assigned an intrinsic, luminous galaxy
shape drawn from equation (7) above, and the intrinsic
position angle of the luminous lens galaxy is taken to be the position angle of 
its dark matter halo.  A source galaxy is placed randomly along the horizontal
axis of the coordinate system, such that it lies within a distance
of $\pm 30''$ of the elliptical lens.  The source galaxy is assigned
a fixed redshift, $z_s = 0.6$.

Next, a circle of radius $60''$, centred on the elliptical lens, is
populated with additional galaxy lenses whose haloes are taken to be 
singular isothermal spheres (SIS).  The number density of the SIS lenses
is matched to the observed number density of galaxies with apparent 
magnitudes $18 \le I_{AB} \le 20$ in the BTC40 data, and they are
assigned random locations within the field.
Each SIS lens is assigned a redshift based upon its apparent magnitude, again
drawn from our adopted probability distribution. 

The elliptical lens is assigned a velocity dispersion, $\sigma_v$, and
truncation radius, $x_t$.  For simplicity, the SIS lenses are assigned
a velocity dispersion equal to the velocity dispersion that is
assigned to the elliptical lens.  Having assigned positions, redshifts,
and gravitational potentials to all of the lenses, then, the net shear
experienced by the source is computed as the sum of the individual shears due
to all lenses (elliptical and SIS) with redshifts $z_l < 0.6$.  In addition,
the final image shape of the lens galaxy residing within the elliptical 
dark matter halo is computed using the net shear due to all foreground
SIS lenses (i.e., SIS lenses with $z_l < z_{el}$).

The above procedure is repeated 20 million times for a fixed axis ratio, $f$, of
the halo of the elliptical lens, fixed values of $\sigma_v$ and $x_t$, 
and fixed source redshift, $z_s = 0.6$.  For each new Monte Carlo
realisation, a new intrinsic position angle and a new intrinsic image
shape for the luminous galaxy within the elliptical lens halo are
generated.   That is, with each new realisation we randomly ``spin'' 
the elliptical lens and
we assign its luminous galaxy a new intrinsic ellipticity.
In addition, the elliptical lens is assigned a new redshift in 
each new realisation.  Thus,
after many realisations, the redshifts of the
elliptical lenses will span the entire range of redshift space that
was adopted for BTC40 galaxies with $18 \le I_{AB} \le 20$.
For each new Monte Carlo realisation
a new location for the source along the horizontal axis is
generated, and
a new suite of SIS lenses is laid down
(including new redshifts and new locations within
the $60''$ circle).  After the source at $z_s = 0.6$
and the elliptical lens at the origin have been lensed by all 
foreground galaxies,
the net shear for  both the source 
and the luminous galaxy within
the elliptical halo are computed in each individual Monte Carlo
realisation.  The mean tangential shear for the sources,
$\gamma^+$ and $\gamma^-$, is then
computed by taking the 
elliptical lenses as the centres for the 
calculation.

\begin{figure}
\begin{centering}
\epsfig{file=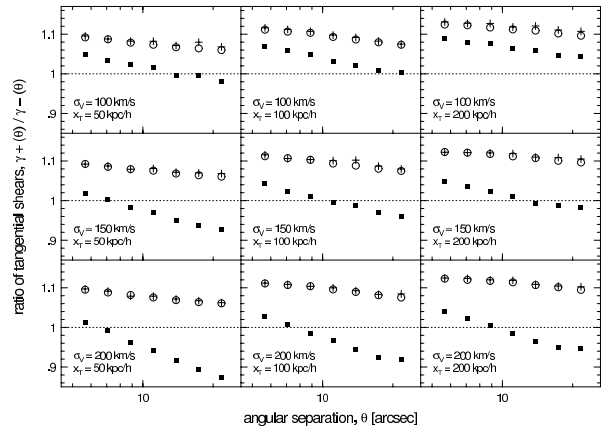,width=2.2in,angle=-90} \\
\end{centering}
\caption{Ratio of mean tangential shears, $\gamma^+ (\theta) /
\gamma^- (\theta)$, for our suite of constrained Monte
Carlo simulations.  Here the halo of the central, elliptical lens galaxy
has $\epsilon_{\rm halo} = 0.1$. The function is measured relative to the
symmetry axes of the central lens galaxy.  
Different panels show results for various values
of the velocity dispersion and truncation radius of the central lens
galaxy.  All source galaxies located within $\pm 45^\circ$ of the symmetry
axes of the central lens galaxy are included in the calculation.  
Error bars are omitted because they are comparable to or smaller than 
the data points.
Circles: Source galaxies have been lensed solely by the central, elliptical
lens. Symmetry axes used in the 
calculation of $\gamma^+ (\theta) / \gamma^- (\theta)$
are the intrinsic symmetry axes of the central
lens.  Crosses: Source galaxies have been lensed by the central, elliptical
lens as well as all foreground SIS lenses.  Symmetry axes used in the calculation
of $\gamma^+ (\theta) / \gamma^- (\theta)$ are the intrinsic symmetry
axes of the central lens.  Squares: ``Observed'' signal.  Source galaxies
have been lensed by the central, elliptical lens as well as all foreground
SIS lenses.  Image of the central, elliptical lens has also been lensed by
all foreground SIS lenses.  Symmetry axes used in the calculation of
$\gamma^+ (\theta) / \gamma^- (\theta)$ are the final, observed symmetry
axes of the central, elliptical lenses after lensing by the foreground SIS lenses. 
}
\end{figure}

Shown in Figures~13-15 is the function 
$\gamma^+ (\theta) / \gamma^-(\theta)$,
obtained by computing the mean tangential shear
for sources located within $\pm 45^\circ$ of the symmetry axes of the
elliptical lens.  Results from a range of halo parameters
($\sigma_v = 100$~km~sec$^{-1}$, 150~km~sec$^{-1}$, 200~km~sec$^{-1}$;
$x_t = 50~h^{-1}$~kpc, $100~h^{-1}$~kpc, $200~h^{-1}$~kpc) are shown
in the different panels.  Figure~13 shows results for central elliptical
lenses in which the ellipticity of the dark matter halo is $\epsilon_{\rm halo} = 0.1$
(corresponding to a projected mass axis ratio $f = 0.82$).  Figure~14 shows
results for central elliptical lenses in which the ellipticity of the dark
matter halo is $\epsilon_{\rm halo} = 0.3$ (corresponding to a projected mass axis 
ratio of $f = 0.54$).  Figure~15 shows results for central elliptical 
lenses in which the ellipticity of the dark matter halo is 
$\epsilon_{\rm halo} = 0.5$ (corresponding to a projected mass axis ratio of $f = 0.33$).

The circles in Figures~13-15 show $\gamma^+ (\theta) / \gamma^- (\theta)$ for
the case that the source galaxies are lensed solely by the central elliptical
lens.  The symmetry axes used for the calculation are the intrinsic symmetry 
axes of the central elliptical lens.  In other words, the circles show
the simplest expected result: all source galaxies are lensed by only one 
foreground galaxy, the lens has a non-spherical dark matter halo,
and the symmetry axes of the lens galaxy are its intrinsic
symmetry axes.  The crosses in Figures~13-15 show $\gamma^+ (\theta) / 
\gamma^- (\theta)$ for the case that the source galaxies are lensed by both
the central elliptical lens, as well as all foreground SIS lenses.  The
symmetry axes used for the calculation are the intrinsic symmetry axes of
the central elliptical lens.  Comparing the crosses to the circles we find
that the introduction of foreground SIS lenses does little to affect the 
ratio of the tangential shears when the intrinsic symmetry axes of the 
central elliptical lens are used for the calculation.

The squares in Figures~13-15 show $\gamma^+ (\theta) / \gamma^- (\theta)$
for the case that the source galaxies are lensed by both the central elliptical
lens, as well as all foreground SIS lenses.  
In addition, the central, elliptical lens has been lensed by all foreground
SIS lenses.
Here the symmetry axes used
for the calculation are the {\it observed} symmetry axes of the central elliptical
lens (i.e., the symmetry axes after lensing by the foreground
SIS lenses).  
From Figures~13-15, the degree to which the observed function, $\gamma^+ (\theta) /
\gamma^- (\theta)$, is suppressed compared to what one would obtain using the
intrinsic symmetry axes of the elliptical lens is a function of the velocity
dispersion that is adopted.
The lower is the velocity dispersion of the lenses, the less the observed function
is suppressed.  This is due to the fact that the frequency and strength of the
multiple weak deflections are lower for lenses with low velocity dispersions than
for lenses with high velocity dispersions (see, e.g., Brainerd 2010).  In contrast,
the ellipticity of the projected dark matter halo of the central elliptical
lens has relatively little effect on the degree to which
the observed function, $\gamma^+ (\theta) / \gamma^- (\theta)$
is suppressed.  (Note that the vertical scales in Figures~13-15 are very different
from each other due to the fact that the more elliptical is the central elliptical
lens, the greater is the anisotropy that it induces.)

\begin{figure}
\begin{centering}
\epsfig{file=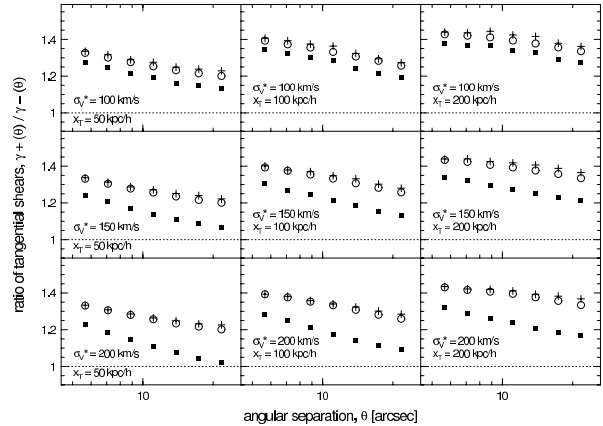,width=2.2in,angle=-90} \\
\end{centering}
\caption{Same as Figure~13 except that here the central, elliptical 
lens has $\epsilon_{\rm halo} = 0.3$.  Note that the vertical scale
differs from that of Figure~13.
}
\end{figure}

\begin{figure}
\begin{centering}
\epsfig{file=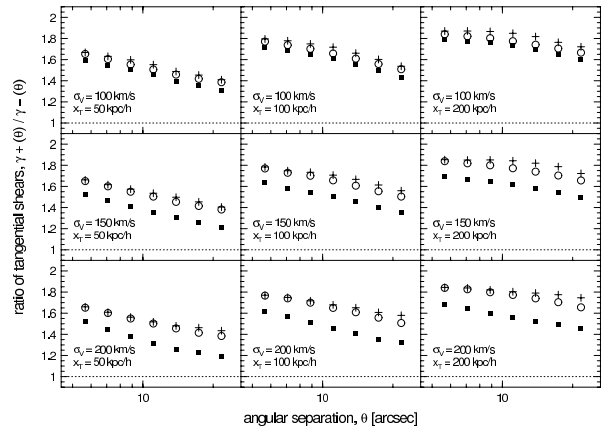,width=2.2in,angle=-90} \\
\end{centering}
\caption{Same as Figure~13 except that here the central, elliptical
lens has $\epsilon_{\rm halo} = 0.5$.  Note that the vertical scale
differs from that of Figure~13.
}
\end{figure}

\section{Judicious Rejection of Lenses and Sources?}

In an attempt to ``inoculate'' one's analysis of galaxy-galaxy lensing
by non-spherical haloes against the above effects,
one might consider simply rejecting
bright centres with small image ellipticities from the calculation of
$\gamma^+ (\theta)$ and $\gamma^- (\theta)$.  That is, one could hope
to avoid the extreme situation where a lens-source pair that is truly in the
``$\gamma^+$'' configuration is swapped to the ``$\gamma^-$'' configuration
due to the ellipticity of the image of the
lens being very small (and, hence, making it
more susceptible to having its symmetry axes altered significantly 
by weak lensing due to foreground galaxies).  Naively, one might hope
that the suppression of the observed function, $\gamma^+ (\theta) /
\gamma^- (\theta)$,
could be eliminated simply by choosing to compute
the mean tangential shear using bright centres whose images are 
highly elliptical. 

In addition, in a search for the signature of anisotropic galaxy-galaxy
lensing, one might be tempted to restrict the analysis to source
galaxies that are very close to the symmetry axes of the 
bright centres that are used to calculate the mean tangential shear.
That is, in all of the analyses above, $\gamma^+(\theta)$ and $\gamma^-(\theta)$
were computed using all sources whose azimuthal coordinates,
$\varphi$, placed them
within $\pm 45^\circ$ of the symmetry axes of the bright centres.  
At fixed angular separation from an elliptical lens, 
the maximal difference in the shear experienced by two sources will, of
course, occur when one source is located along the minor axis of the
lens and the other
is located along the major axis of the lens.  Therefore, one might expect
that if one narrowed the analysis region from $\pm 45^\circ$
to, say, $\pm 25^\circ$ or $\pm 15^\circ$,
it would be easier to detect anisotropic galaxy-galaxy lensing.

Unfortunately, the situation is not that simple in either
of these cases.   
Weak lensing of the bright centres may make their
resulting images either rounder or more elliptical than their intrinsic
image shape (i.e., bottom panel of Figure~12).  Suppose that one chooses
a minimum image ellipticity for the bright centres, and that the computation
of $\gamma^+ (\theta)$ and $\gamma^- (\theta)$ is performed using only those
bright centres with observed ellipticity $\epsilon_{\rm light} > \epsilon_{\rm cut}$.  
Some fraction of the bright centres whose intrinsic ellipticity 
truly exceeds
$\epsilon_{\rm cut}$ will, by weak lensing by foreground galaxies, have
their resulting images made rounder than their intrinsic 
ellipticity.   
As a result, some bright centres with intrinsic ellipticities that are 
larger than $\epsilon_{\rm cut}$ 
will, in fact, be rejected on because their
observed (post-lensing) images have ellipticities smaller than $\epsilon_{\rm cut}$.  
The number of such bright centres that
are affected by this will
vary with the magnitude of the shear that they experience.

In addition to changing the ellipticity, weak lensing of the bright centres
may rotate the orientations of their symmetry axes (i.e., top
panel of Figure~12).  If one simply tries to narrow one's analysis
region relative to the symmetry axes of the 
bright centres, a problem will occur if the bright centres have been
weakly lensed.  Any rotation of the symmetry axes of the bright centres causes
the analysis region that one truly desires (i.e., the region that brackets
the directions of the major and minor axes of the projected halo mass)
to be rotated with respect to the analysis region that one must actually
use in practice (i.e., the region that brackets the directions of
the observed major and minor axes of the image of the bright centre).  Therefore,
narrowing the analysis region may actually increase the discrepancy between the
observed function, $\gamma^+ (\theta) / \gamma^- (\theta)$, and the function
that one would measure if the intrinsic symmetry axes of the bright centres
were known.  

\begin{figure}
\begin{centering}
\epsfig{file=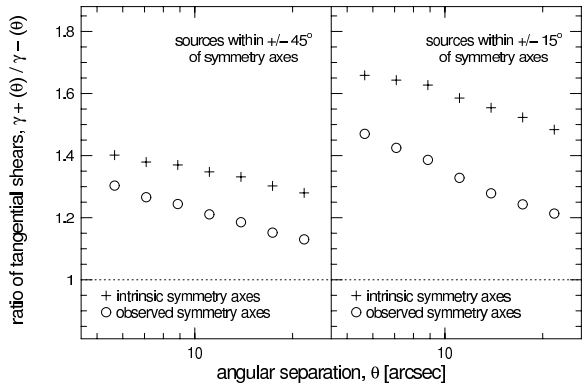,width=3in,angle=0} \\
\end{centering}
\caption{
Effect of narrowing the analysis region on the ratio of mean tangential
shears.  Here a fiducial elliptical lens with $\sigma_v = 150$~km~sec$^{-1}$,
$x_t = 100~h^{-1}$~kpc, and projected halo ellipticity $\epsilon_{\rm halo} = 0.3$ has
been adopted.  
Circles: ``Observed'' signal.  Sources have been lensed by the central,
elliptical lens and all foreground SIS lenses.  The central, elliptical lens
has also been lenses by all foreground SIS lenses and its observed symmetry
axes are used for the calculation.  Crosses:  Sources have been lensed
by the central, elliptical lens and all foreground SIS lenses.  The intrinsic
(unlensed) symmetry axes of the central lens are used for the calculation.
{\it Left:} All sources within $\pm 45^\circ$ of the 
lens symmetry axes are used in
the calculations.  {\it Right:} Only sources within $\pm 15^\circ$ of the lens
symmetry axes are used in the calculations.  Narrowing the analysis region
increases the degree of anisotropy in the galaxy-galaxy lensing signal, but it
also increases the disparity between the observed signal and the one that would
be measured if the intrinsic symmetry axes of the lenses were known.
}
\end{figure}

In this section we adopt a fiducial elliptical lens with velocity dispersion
$\sigma_v = 150$~km~sec$^{-1}$, truncation radius $x_t = 100~h^{-1}$~kpc, and
projected halo ellipticity $\epsilon = 0.3$, and we construct Monte Carlo simulations
that are identical to the Monte Carlo simulations in Section 6
(i.e., the simulations used to obtain the central panel of Figure~14).
Shown in Figure~16 is the effect of narrowing the analysis region when computing
the ratio of the mean tangential shears.  The left panel of Figure~16 shows
the observed and intrinsic functions, $\gamma+^ (\theta) / \gamma^- (\theta)$,
when all sources within $\pm 45^\circ$ of the symmetry axes of the central,
elliptical lens are used for the calculations.  The right
panel of Figure~16 shows the same functions as the left panel, but here only
sources that are within $\pm 15^\circ$ of the symmetry axes of the central,
elliptical lens are used for the calculations.  From the right panel of Figure~16,
it is clear that narrowing the analysis region (i.e., using only sources that
are very close to the symmetry axes) increases the degree of anisotropy in
the galaxy-galaxy lensing signal.  However, narrowing the analysis region also
increases the disparity between the observed function, $\gamma^+ (\theta) /
\gamma^- (\theta)$, and the function that would be measured if the intrinsic
symmetry axes of the central, elliptical lenses were known.  

Figure~17 shows the effect of rejecting bright centres whose images (post-lensing)
are very round.  
All sources within $\pm 45^\circ$ of the symmetry axes of the central, elliptical
lens are used in the calculation.
Here circles show the observed function, $\gamma^+ (\theta) /
\gamma^- (\theta)$, and crosses show the function that one would obtain if the 
intrinsic symmetry axes of the central elliptical lens were known.  In the case
of the circles and crosses, no constraint on the ellipticity of the lens image
is imposed.  Triangles in Figure~17 show the observed function,
$\gamma^+ (\theta) / \gamma^- (\theta)$, where the ratio has been computed using
the observed symmetry axes of central, elliptical galaxies whose images (post-lensing)
have ellipticities $\epsilon_{\rm light} > 0.3$.  From this figure, then, rejection of lenses
with image ellipticities $\epsilon_{\rm light} < 0.3$ increases the observed function, 
$\gamma^+ (\theta) / \gamma^- (\theta)$, only slightly.  In particular, rejection of
the lenses with the roundest images does not allow one to recover the function
that one would measure if the intrinsic symmetry axes of the central, elliptical
lenses were known.

\begin{figure}
\begin{centering}
\epsfig{file=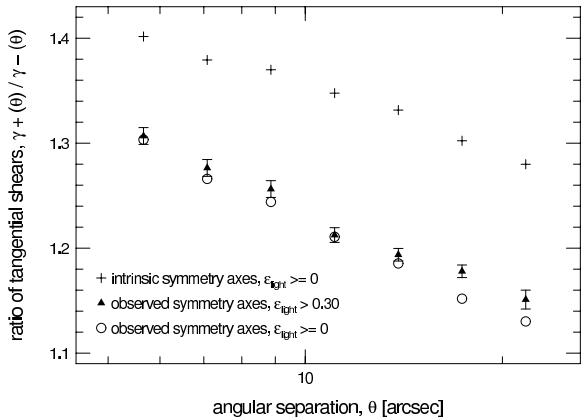,width=3in,angle=0} \\
\end{centering}
\caption{Effect of rejection of round lenses on the ratio of the mean tangential
shears.  Here all sources within $\pm 45^\circ$ of the symmetry axes of the
central, elliptical lens are used.  The central, elliptical lens
has halo parameters $\sigma_v = 150$~km~sec$^{-1}$,
$x_t = 100~h^{-1}$~kpc, and projected ellipticity $\epsilon_{\rm halo} = 0.3$.
Error bars are omitted when they are comparable to or smaller than the data
points.
Circles: ``Observed'' signal.  Sources have been lensed by the central,
elliptical lens and all foreground SIS lenses.  The central, elliptical lens
has also been lenses by all foreground SIS lenses and its observed symmetry
axes are used for the calculation.  No constraint has
been placed on the ellipticity of the image of the central, elliptical
galaxy.  Crosses:  Sources have been lensed
by the central, elliptical lens and all foreground SIS lenses.  The intrinsic
(unlensed) symmetry axes of the central lens are used for the calculation.
No constraint has been placed on the ellipticity of the image of the central, elliptical
galaxy.  Triangles: ``Observed'' signal, computed using only those central,
elliptical lenses whose observed (post-lensing) image ellipticity is
$\epsilon_{\rm light} > 0.3$.  Rejection of central, elliptical galaxies with
very round images does little to affect the discrepancy between the observed
function and the one that would be obtained if the intrinsic symmetry axes of
the central, elliptical lenses were known.
}
\end{figure}

\section{Summary and Conclusions}

We have investigated the theory of
galaxy-galaxy lensing by non-spherical dark matter
haloes, which should give rise to an anisotropy in the tangential shear
experienced by distant source galaxies.  
If each distant source is lensed by 
only one foreground elliptical lens, and if the observed
symmetry axes
of the elliptical lens correspond to the intrinsic symmetry axes of its
projected dark matter halo, one would expect the signature of anisotropic 
galaxy-galaxy lensing to manifest as $\gamma^+ (\theta) > \gamma^- (\theta)$
over a wide range of angular scales.
Here $\gamma^+ (\theta)$ is the angular dependence of the mean tangential
shear experienced by sources whose azimuthal coordinates place them close
to the major axis of the lens, and $\gamma^- (\theta)$ is the angular
dependence of the mean tangential shear experienced by sources whose
azimuthal coordinates place them close to the minor axis of the lens.

Using an observational data set (observed coordinates and $I$-band
apparent magnitudes) as a framework for a set of Monte Carlo simulations,
we have demonstrated that the actual signature that one should expect
to observe for anisotropic galaxy-galaxy lensing is far from the 
above idealised case.  
Because galaxies are broadly distributed in
redshift space, it is common for a distant source galaxy located 
at redshift $z_s$ to be 
lensed by another galaxy located at redshift $z_{l1} < z_{s}$.  In turn,
this original lens-source pair may then be lensed by yet another galaxy (or galaxies)
located at redshift $z_{l2} < z_{l1}$.  Such instances of ``multiple
deflections'' cause the observed signature of anisotropic galaxy-galaxy
lensing to deviate from the expected signature.  The degree to which the observed
signature of galaxy-galaxy lensing deviates from the expected signature is
a strong function of the characteristic velocity dispersion of the haloes of
$L^\ast$ galaxies.  In the case of low characteristic velocity dispersions,
$\sigma_v^\ast = 100$~km~sec$^{-1}$, the observed ratio of mean tangential
shears, $\gamma^+ (\theta) / \gamma^- (\theta)$, exceeds a value of unity 
on all scales $\theta < 100''$ and is only slightly lower than the function
one would obtain if the intrinsic symmetry axes of the foreground galaxies were used to
perform the calculation.  In the case of moderate velocity dispersions,
$\sigma_v^\ast = 150$~km~sec$^{-1}$, the observed ratio of mean tangential shears
shows little to no anisotropy on scales $\theta > 20''$.  In the case of
high velocity dispersions, $\sigma_v^\ast = 200$~km~sec$^{-1}$, the observed
function is actually reversed from the expected function (i.e., $\gamma^+ (\theta) <
\gamma^- (\theta)$) on scales $20'' < \theta < 70''$, and is consistent with
no anisotropy on scales $70'' < \theta < 120''$.

In summary, 
our simulations show that if one observes $\gamma^+ (\theta) = 
\gamma^- (\theta)$ in a large galaxy-galaxy lensing data set,
the observation cannot be simply interpreted as proof that the haloes of the
lens galaxies are spherically-symmetric.  That is, although the measured
signal appears to be isotropic, it is entirely possible that
anisotropic galaxy-galaxy lensing
by non-spherical haloes may have taken place. Further,  
our simulations show that if one observes $\gamma^+ (\theta) <
\gamma^- (\theta)$ in a large galaxy-galaxy lensing 
data set, the observation cannot be simply interpreted as
proof that mass and light
are ``anti-aligned'' in the lens galaxies.  That is, although the
measured signal appears to be reversed from the expected
signal, the reversal may occur when mass and light are, in fact, perfectly
aligned within the lens galaxies.

The primary reason that the observed signature of anisotropic
galaxy-galaxy lensing differs from the expected signature
is that the foreground
galaxies that are used as centres to compute the mean tangential shear
have, themselves, been weakly lensed.  The expectation that
$\gamma^+ (\theta)$ will exceed $\gamma^- (\theta)$ over a wide
range of angular scales is based upon a picture 
in which the observed symmetry axes of the lenses are identical to the
intrinsic symmetry axes of their projected dark matter haloes.  However,
when one computes $\gamma^+ (\theta)$ and $\gamma^- (\theta)$ in
an observational data set, one cannot 
directly view the intrinsic symmetry
axes of the bright, central galaxies.  Instead, one is forced to use 
their observed 
symmetry axes and, in general, these will differ from the intrinsic
symmetry axes.

Our simulations show that, even in the limit of multiple deflections
being experienced by the distant source galaxies, if one could use
the intrinsic symmetry axes of the lenses to define the 
geometry of the problem, one would expect to observe $\gamma^+ (\theta)
 > \gamma^- (\theta)$.   That is, multiple deflections experienced by
the source galaxies have little effect on the intrinsic 
signature of anisotropic
galaxy-galaxy lensing by non-spherical haloes.  However, weak lensing of 
the bright, central foreground galaxies causes their observed
symmetry axes (which are used to define the geometry for the calculation
of $\gamma^+ (\theta)$ and $\gamma^- (\theta)$) to 
differ from their intrinsic symmetry axes (i.e., the unlensed symmetry axes,
which define the geometry for the actual lensing of
the distant galaxies).  It is this change in the symmetry axes of the 
bright, foreground galaxies that gives rise to the suppression of the
observed function, $\gamma^+ (\theta) / \gamma^- (\theta)$, compared to 
the function that would be obtained if the intrinsic symmetry axes were
used for the calculation.  The effects of weak lensing of the bright,
foreground galaxies on an observation of $\gamma^+ (\theta) / \gamma^- (\theta)$
cannot be eliminated simply by rejecting foreground galaxies with very
small image ellipticities, or by using sources that are particularly
close to the observed symmetry axes of the foreground galaxies.

We conclude, therefore, that
in order to properly interpret any observed galaxy-galaxy 
lensing signal (be it isotropic or anisotropic), it
is vital that full, multiple-deflection Monte Carlo simulations be used.
Especially important is accounting for the fact that the images of the
bright, foreground centres are likely to have been weakly lensed.  If the
effects of multiple deflections are not taken into account when 
interpreting an observed galaxy-galaxy lensing signal, there is a
high probability that incorrect conclusions will be drawn
about the nature of the haloes surrounding the lens galaxies.

\section*{Acknowledgments}
It is a pleasure to thank the BTC40 survey team, particularly Emilio Falco,
Chris Kochanek, Malcolm Smith and Richard Green, for allowing us to use
their data.
Support from the National Science Foundation under NSF contracts
AST-0406844 and AST-0708468 is gratefully acknowledged.

\label{lastpage}

\end{document}